\documentstyle[psfig]{lamuphys}
\makeatletter
\let\chapter\hid@chapter
\makeatother

\begin{document}

\ifx\href\undefined
\else\errmessage{Don't use HyperTeX}
\fi

\pagenumbering{arabic}
\title{The Mg--$\sigma$ Relation of Elliptical Galaxies at Various Redshifts}

\author{Bodo L. Ziegler}

\institute{Universit\"atssternwarte, Scheinerstra\ss e 1,
D--81679 M\"unchen, Germany}

\maketitle

\begin{abstract}
The correlation between the Mg absorption index and the velocity dispersion
($\sigma$) of local elliptical galaxies is very tight. Because the Mg
absorption depends on both metallicity and age of the underlying stellar
population the observed Mg--$\sigma$ relation constrains the possible
variation in metallicity and age for a given velocity dispersion. For a time
interval with no change in metallicity any variation of the Mg index is caused
only by the aging of the stars. 

We have measured the Mg absorption and velocity dispersion of ellipticals in
three clusters at a redshift of $z=0.37$ and established their Mg--$\sigma$
relation. For any given $\sigma$, the measured Mg absorption is weaker than
the mean value for local ellipticals. Since the evolution of bright cluster
ellipticals between $z=0.4$ and today is most probably only `passive' this
reduction in Mg can be attributed solely to the younger age of the stellar
population. The small weakening of the Mg absorption of the distant galaxies
compared to the local values implies that most of the stars in cluster
ellipticals must have formed at high redshifts ($z_f>2\ldots 4$).

The Mg--$\sigma$ test is a very robust method to investigate the evolution
of elliptical galaxies and has several advantages over traditional methods
using luminosities. A remaining problem is the aperture correction necessary
to calibrate observations of galaxies at different distances. Here, we show
that our general conclusions about the epoch of formation still hold when
aperture corrections are calculated assuming a dependence of the radial
gradient of $\sigma$ on the galaxy's effective radius rather than assuming
no dependence as was done in all previous studies.
\end{abstract}

\section{The Local Mg--$\sigma$ Relation}
It is well known that all dynamically hot galaxies in the local universe
follow the same linear relationship between the Mg absorption around
$\lambda_0\approx5170$~\AA\ and the internal velocity dispersion $\sigma$
(\cite{DLBDFTW87,BBF93}). Although the galaxies span a wide range in Mg and
$\sigma$, the Mg--$\sigma$ relation is very tight. A sample of luminous Coma
ellipticals ($\lg\sigma\ge2.3$), e. g., with Mg$_2\in[0.25,0.36]$~mag has a
standard deviation from the linear fit of only $\sigma_{\rm int}=0.011$~mag.
Mg$_2$ as defined in the Lick system (\cite{FFBG85}) comprises mainly the
molecular absorption of MgH. Because the measurement of this index in
redshifted galaxies is very noisy we use the atomic Mg$_b$ index instead. A
linear transformation from Mg$_2$ to Mg$_b$ enables us to still use the
7~Samurai sample of Coma and Virgo ellipticals as the comparison at zero
redshift. Both from observational (\cite{Gonza93}) and theoretical data
(stellar population synthesis of \cite{Worth94}) we derived consistently:
Mg$_b/{\rm \AA}=14.9\pm0.5\cdot{\rm Mg}_2/{\rm mag}$. In figure~\ref{mgbs}
small circles present the Mg$_b$--$\sigma$ relation of the local comparison
sample. A principal component analysis yields as best fit:
\begin{equation}
\label{gl_mgbs}
\mbox{Mg}_b = 2.7 \lg \sigma_0 - 1.65
\end{equation}

The dependence of Mg$_b$ on age and metallicity can be explored using
stellar population synthesis models. From Worthey's 1994 models  we
derived the following formula that holds for ages $t>3$~Gyrs and
metallicities $-2<\lg Z/Z_\odot<+0.25$:
\begin{equation}
\label{gl_mgbtZ}
\lg \mbox{Mg}_b = 0.20 \lg t + 0.31 \lg Z/Z_\odot + 0.37
\end{equation}
This formula allows to determine the maximum variation of both relative age
and relative metallicity as it is constrained by the very tightness of the
Mg$_b$--$\sigma$ relation. For a given velocity dispersion $\sigma$ (with
$\lg\sigma\ge2.3$) and zero variation in metallicity or age, resp., we find:
\begin{equation}
\label{gl_tZdisp}
\Delta t/t < 0.17 \quad \mbox{and} \quad \Delta Z/Z < 0.11
\end{equation}
This narrow constraint on the age spread of cluster ellipticals implies that
they did not form continuously at the same rate but that there was a rather
short formation epoch of these galaxies. If, e.g., the majority of
ellipticals were formed 12~Gyrs ago, then the scatter in age would be about
2~Gyrs. However the formation epoch itself can only be determined by comparing
the relative ages of galaxies at zero and a significantly higher redshift.

\section{The Mg$_b$--$\sigma$ Relation at Redshift $z=0.37$}
As a first step we have established the Mg$_b$--$\sigma$ relation of
elliptical galaxies in three clusters at a redshift of $z=0.37$ ({\it
Abell~370, CL~0949$+$44} and {\it MS~1512$+$36}) (\cite{BZB96}). The spectra
were taken with the 3.5m~telescope at the Calar Alto observatory and the
3.6m and the NTT at ESO with total integration times per galaxy on the order
of 8 hours. Very careful data reduction had been applied because in the
relevant wavelength range ($\lambda({\rm Mg}_b,z=0.37)\approx7080$~\AA) the
spectra are heavily contaminated by strong night sky emission lines and
telluric absorption bands (\cite{ZB97}).

\begin{figure}[hbt]
\begin{center} \parbox{10cm}{ 
\psfig{figure=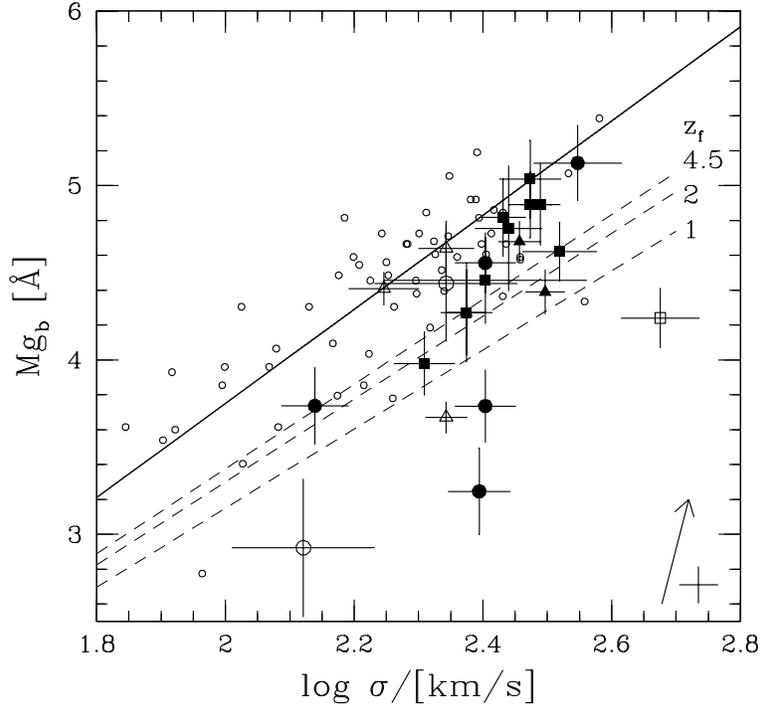,width=10cm}}
\caption[a]{\label{mgbs} 
Mg$_b-\sigma$ pairs at $z=0.37$ (big symbols with
errorbars) compared to the local Mg$_b$--$\sigma$ relation
(small circles: Coma and Virgo elipticals, typical
errorbar in lower right corner). Arrow: aperture correction applied. Dashed
lines: expected Mg$_b$--$\sigma$ relations at $z=0.37$ for $z_f=1,2,4.5$ and 
$H_0=50, q_0=0.5$. }
\end{center} 
\end{figure}
Figure~\ref{mgbs} presents the datapoints for the distant galaxies together
with the local Mg$_b$--$\sigma$ relation. The first thing to note is that
all distant ellipticals have lower Mg$_b$ line strengths than the local mean
value at the same velocity dispersion. This is clear evidence for evolution
of the stellar populations of elliptical galaxies between $z=0.37$ and now.
On the other hand the reduction in Mg$_b$ is very weak, on average
$<\!\!\Delta\mbox{Mg}_b\!\!>=-0.37\pm0.08$~\AA. This can be reconciled with
current stellar population models only if there was virtually no new star
formation in today's cluster ellipticals since $z=0.37$ but only `passive'
evolution of the aging stars. Therefore, the metallicity did most probably
not change at all and the reduction in Mg$_b$ can be fully attributed to the
younger age of the distant galaxies. Setting $\Delta Z=0$,
equation~\ref{gl_mgbtZ} together with equation~\ref{gl_mgbs} can be
transformed to deduce theoretical curves of the Mg$_b$--$\sigma$ relation at
the observed redshift $z=0.37$:
\begin{equation}
\label{gl_mgbt}
\frac{\mbox{Mg}_{\,b\,}(z\!=\!0)}{\mbox{Mg}_{\,b\,}(z)} = 
\left( \frac{\mbox{age}\,(z\!=\!0)}{\mbox{age}\,(z)} \right)^{0.20}
\end{equation}
The age of an object depends mainly on its redshift of formation $z_f$ and
less on the cosmology ($H_0, q_0, \Lambda$). In figure~\ref{mgbs}, the
dashed lines correspond to the expected Mg$_b$--$\sigma$ relation at
$z=0.37$, if $z_f$ were 1, 2 or 4.5, resp.. The small weakening of Mg$_b$ to
a look--back time of $\approx5$~Gyrs constrains the age of the stellar
populations so that the majority of the stars in elliptical cluster galaxies
were formed at high redshifts $z_f\geq2$. For the most luminous ellipticals,
where the reduction in Mg$_b$ is even lower, $z_f$ could be as high as 4.
This imposes great problems on current theories on the structure formation
using Cold Dark Matter models, because contrary to these models smaller
objects seem to be younger than big ones and formed at later times. But a
large dynamically relaxed stellar system most probably did not form within
only 1~Gyr after the Big Bang as is fixed by $z_f=4$ in all reasonable
cosmologies. A way out of this paradigm is a model in which the stars indeed
existed already at $z\geq4$ in a common gravitational potential but the
galaxies were formed only at much later times through the collapse of these
dark matter halos and/or through the dissipationless merging of smaller
halos. The smaller elliptical galaxies could either have a much more
extended formation epoch or experience events during their evolution which
add new populations of stars and therefore lower their mean age.

The evolution of the stellar population as measured by the Mg$_b$--$\sigma$
test can be transformed into a change in luminosity with the help of
population synthesis models. From the Worthey and
Bruzual \& Charlot (1997) models we found consistently the following linear
relation between the reduction in Mg$_b$ and the increase in the $B$--band
luminosity valid for ages greater than 1.5~Gyrs and metallicities between
half and twice solar:
\begin{equation}
\label{gl_mgbmb}
\Delta \mbox{M}_B \,\mbox{[mag]} \quad \approx \quad  1.35 \pm 0.1
\,\Delta \mbox{Mg}_b \,[\mbox{\AA}]
\end{equation}
Thus, the distant cluster ellipticals are on average brighter in the
$B$--band by $<\!\!\Delta \mbox{M}_B\!\!>=-0.50\pm0.11\mbox{[mag]}$
than local ones. This is quantitatively consistent with the brightening
obtained via the Faber--Jackson relation. 

The Mg$_b$--$\sigma$ test is, therefore, a powerful and reliable tool to
investigate the evolution of elliptical galaxies. Over all other methods
using luminosities or colors it has the advantage to be free of problems
like foreground and internal extinction and K--corrections. It also depends
only very weak on the initial mass function. The accurate determination of
the luminosity evolution by this method, therefore, makes it possible for
the first time to calibrate elliptical cluster galaxies as standard candles.
Together with the fundamental plane relations the cosmological parameter
$q_0$ can be significantly constrained (see the contribution by Bender
et al., this conference or 1997).

\begin{figure}[hbt]
\begin{center} \parbox{10cm}{ 
\psfig{figure=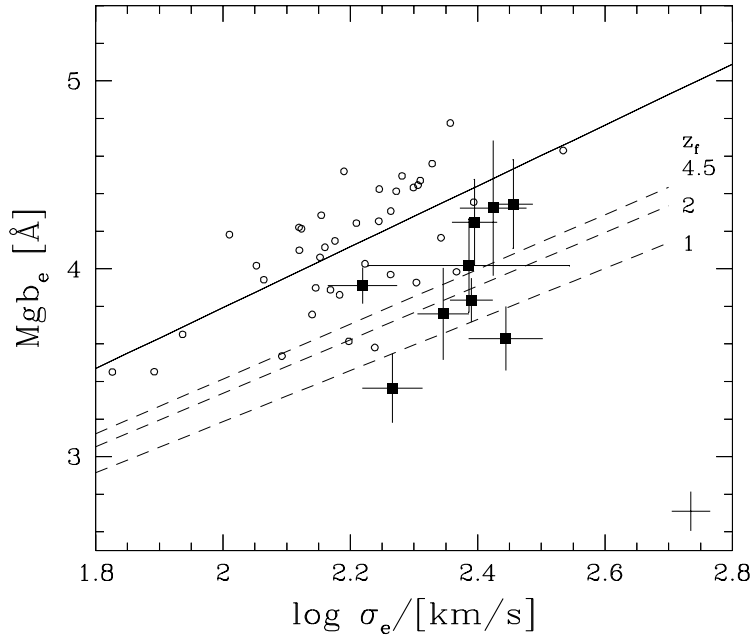,width=10cm}}
\caption[a]{\label{mgbsre} 
Same diagram as figure~\ref{mgbs} but with
Mg$_b$ and $\sigma$ extrapolated to integrated mean values within $r_e$. }
\end{center} 
\end{figure}

The weak part of the Mg$_b$--$\sigma$ test is the aperture correction
necessary to calibrate observations of galaxies at different distances. Even
if our aperture size were the same as was used for the observations of the
local comparison sample we would average our measured values over a much
greater fraction of the galaxies. Indeed, the 7~Samurai determined central
values whereas we integrate out to more than the effective radius ($r_e$).
Because both Mg$_b$ and $\sigma$ have radial gradients an aperture
correction has to be applied. From spectroscopic observations of ellipticals
with sufficient $S/N$ out to several $r_e$ (Saglia {\it et al.},
unpublished) we found the following gradients:
\begin{equation}
\label{gl_mgbgrad}
{\rm Mg}_b = -0.87 \lg (r/r_e) + c
\end{equation}
\begin{equation}
\label{gl_siggrad}
\lg \sigma (r) = -0.11 \, (r/r_e)^{3/4} + c
\end{equation}
The gradient of $\lg \sigma$ depends on the effective radius in a non
logarithmic manner unlike those used in previous studies but which were
determined from data with a much lower radial extent (e.g. \cite{JFK95}) and
falls off steeply at $r_e$. Because we could not determine the effective
radii of all the observed distant ellipticals from our ground--based
photometry (\cite{Ziegler97}) we chose a representative value and applied
the same aperture correction for all galaxies in figure~\ref{mgbs}.
Recently, we have been able to accurately measure $r_e$ for 9 galaxies from
our HST images of {\it Abell~370} and {\it MS~1512$+$36} and, thus, apply
individual aperture corrections. Figure~\ref{mgbsre} presents the same
diagram as figure~\ref{mgbs} but with Mg$_b$ and $\sigma$ extrapolated to
integrated mean values within $r_e$. Although individual galaxies changed
their position in the new diagram our general conclusions about the age
($z_f$) and evolution are still valid with an average reduction in Mg$_b$ of
$<\!\!\Delta\mbox{Mg}_b\!\!>=-0.34\pm0.19$~\AA.

\section{The Mg$_b$--$\sigma$ Relation at Higher Redshifts}
Going to higher redshifts we expect to find a more prominent evolution in
Mg$_b$ because the stellar population gets still younger and the formation
epoch is approached. The higher amount of evolution and the wider range in
look--back time should enable us to reduce further the errors in the
calibration of elliptical cluster galaxies as standard candles and allow us
to even better constrain the value of the cosmological parameter $q_0$. But
if future data at high redshifts would show evolution allowing
merging/accretion events in addition to passive evolution, then possible
variations on the velocity dispersions must be taken into account.

Clearly, the spectroscopy of very faint objects will be a major challenge.
Observing at higher redshifts means that the Mg$_b$ absorption line moves
more and more into the red wavelength range with a strongly increasing sky
contamination. The next favourable redshift bins where there is a slight
depression in the sky emission are: $0.575<z<0.583$ and $0.754<z<0.780$. The
decrease in brightness of the galaxies due to their greater distance
(1\ldots 2 mags) might just be compensated by their evolutionary
brightening. In order to get spectra with $S/N$ as good as we have obtained
at $z=0.37$ the suggested observations can only be done with a big enough
telescope like the VLT together with a spectrograph of superb efficiency
like FORS. Exposure times would be on the order of 1.5\ldots 2 hours.
Multi--object spectroscopy will be the appropriate method if the apertures
are large enough to collect a sufficiently large area of the sky around each
object in order to be able to cope with the sky subtraction.

\section{Conclusions}
Determining the Mg$_b$--$\sigma$ relations at various redshifts is a
powerful and robust method to measure the evolution of elliptical galaxies.
From the first application at a redshift of $z=0.37$ we could
already make two firm conclusions: first, the stellar population of elliptical
cluster galaxies evolves predominantly passive between this redshift and
today and second, the epoch of formation of the stars that mainly make up the
ellipticals today is at high redshifts. The evolution in Mg$_b$ can be
reliably transformed into an increase of the $B$--magnitude, thus, allowing
the calibration of elliptical cluster galaxies as standard candles.

\bigskip

{\bf Acknowledgements:} This work is a collaboration with Drs. Bender,
Belloni, Bruzual and Saglia and was supported by the
``Sonderforschungsbereich 375--95 f\"ur Astro--Teilchenphysik der Deutschen
Forschungsgemeinschaft'' and by DARA grant 50 OR 9608 5.

%
%
%

\end{document}